\documentclass[10pt]{article}
\usepackage{epsfig}
\begin{document}
\begin{center}
\LARGE\bf{Gamma Rays from Near-Solar WIMP Annihiliations}\\
\end{center}
\paragraph*{}
\begin{center}
\large\bf{D.W. Hooper}\\
\end{center}
\begin{center}
\large{Phenomenology Institute}\\
\large{University of Wisconsin, Madison}\\
\end{center}
\vspace{1cm}

\LARGE{Abstract}\
\setlength{\parindent}{0pt}
\normalsize{}
\paragraph*{}
If weakly-interacting massive particles (WIMPs) make up the galactic dark matter halo, high densities of such particles would exist near massive bodies, such as the sun.  The resulting annihilations are believed to produce neutrinos, gamma rays and other decay products.  If any significant fraction of such annihiliations occur outside of the sun itself, gamma rays may be observable by detectors on Earth.  
\paragraph*{}
This paper describes a calculation to determine the fraction of solar-captured WIMP annihiliations which occur outside of the sun.  The implications of this result on experimental searches is also discussed.  We find that the prospect of detecting gamma rays from near-solar WIMP annihilations is very poor.
\paragraph*{}
 
\LARGE{I. Introduction} \
\setlength{\parindent}{0pt}

\normalsize{}
\paragraph*{}
There has been a great deal of evidence mounting recently which supports the existence of cold, weakly-interacting, particle dark matter.  It is believed that such matter is the primary constituent of the galactic dark matter halo density.  Although the mass of such a particle is model dependent, it is generally in the range of 10's of GeV to several TeV.  Such particles interact with massive bodies such as the Sun and Earth and become gravitationally trapped due to the loss of kinetic energy during the interaction.  A high density of such particles then builds up in the center of these bodies until the annihilation rate of WIMPs equals the capture rate.  
\paragraph*{}
Following the initial interaction between a WIMP and the Sun (or other massive body), a WIMP, if captured, will assume an elliptical orbit which passes through but does not generally remain within the Sun during the entire orbit.  It often takes many interactions between the WIMP and sun before total internal capture occurs.  If the WIMP becomes totally captured within the Sun, it will eventually annihilate and will be observable only if it's annihilation modes include neutrinos (other annihilation products will not penetrate the Sun).  If the WIMP happens to annihilate with another WIMP outside of the Sun (prior to total capture), other annihilation products can be observed, including gamma rays.
\paragraph*{}
Extensive calculations have been made to predict the number of neutrinos produced in the Sun from WIMP annihilations.$^{1,2}$  The only additional information needed to predict a similar result for gamma rays is the fraction of WIMP annihilations that occur outside of the Sun and the comparison of the branching ratios of WIMPs into neutrinos versus photons.
\paragraph*{}

\LARGE{II. Calculations} \
\setlength{\parindent}{0pt}

\normalsize{}
\paragraph*{}
Using a Boltzman velocity distribution with rms velocity 270 km/sec for the WIMP halo density, the temperature of the population is given by:

\begin{equation}
T=\frac{2 m v_{rms}^2}{\pi k}
\end{equation}

where m is the WIMP mass.  Then a random velocity can be assigned to a WIMP using Boltzman statistics:

\begin{equation}
v=\sqrt{\frac{2 E}{m}}=\sqrt{\frac{-2 k T \ln x}{m}}
\end{equation}

where x is a random number between one and zero selected by the monte carlo.  Assuming the WIMP travels through, and interacts with, the Sun, it first gains 1385 km/sec from solar gravitational potential and then loses an average velocity of $^3$:

\begin{equation}
\delta v = 2 v m_{p}/m_{WIMP} 
\end{equation}

If the resulting velocity, as calculated at infinity, is less than zero, the WIMP is captured.  If it is greater than -619 km/sec, the solar surface escape velocity, some of the WIMPs orbit will be outside of the Sun.
\paragraph*{}
To simplify the problem, this calculation will assume that the WIMP takes a direct trajectory into the center of the Sun (zero angular momentum).  This is a reasonable approximation since the majority of WIMPs captured will travel generally through the center of the Sun rather than through an edge.  This is a result of the greater amount of solar material passed through in a direct trajectory.  
\paragraph*{}
From the new velocity, simple conservation of energy can determine the semi-major axis of the one-dimensional WIMP orbit:

\begin{equation}
\frac{1}{2}v_{surface}^{2}-\frac{G M_{Sun}}{R_{Sun}}=0-\frac{G M_{Sun}}{a}
\end{equation}
\begin{equation}
a=\lbrack \frac{1}{R_{Sun}}-\frac{v_{surface}^{2}}{2 G M_{Sun}} \rbrack ^{-1}
\end{equation}

where a is the semi-major axis of the new WIMP orbit.  From this and Kepler's laws, the orbital period is easily calculated: 

\begin{equation}
Period=\frac{2 \pi a^{3/2}}{G^{1/2} M_{Sun}^{1/2}}
\end{equation}

  After a number of orbits determined by the probability for a passing WIMP to interact with the Sun, the WIMP will further lose energy and a new semi-major axis and period will be calculated.  This is continued until the new semi-major axis is less than the solar radius and total solar capture occurs.  The total distance and total time spent outside of the Sun by the WIMP are calculated by suming the semi-major axes and periods including a factor reflecting the fraction of the orbit spent outside the Sun which is generally close to one.
\paragraph*{}
The probability of a WIMP annihilating with another near-solar WIMP is calculated by integrating the product of the total distance traveled outside of the Sun prior to total solar capture, the WIMP-WIMP cross section and the number of density of WIMPs in the region.  The WIMP-WIMP cross section can be approximated reasonably well and is generally on the order of $\ 10^{-35} cm^{2}$ for a WIMP with velocity of a few hundred kilometers per second.$^4$  The number density of near solar WIMPs is calculated by iteration in the monte carlo.  It is approximated by a flat distribution extending to the mean semi-major axis of a near-solar WIMP orbit (usually around 2 solar radii).  To determine if this approximation was valid, a distribution was calculated.  It revealed a strong suppression at large distance from the Sun.  The majority of WIMP interactions should occur within a few solar radii.  The density distribution was normalized by the solar WIMP capture rate which is generally around $\ 10^{21}$ WIMPs per second and varies inversely with WIMP mass.$^5$ 
\paragraph*{}

\LARGE{III. Results and Discussion} \
\setlength{\parindent}{0pt}

\normalsize{}
\paragraph*{}
To begin calculating prospective event rates from near-solar WIMP annihilations, some assumptions are needed.  The WIMP-WIMP and WIMP-proton cross sections are approximated by the standard weak coupling with appropriate suppressions.  A value of $6 \times 10^{-42} cm^2$ was used for WIMP-proton interactions and a range of values from $2 \times 10^{-36}$ to $2 \times 10^{-34} cm^2$ were used for WIMP-WIMP interactions.  Three values of the WIMP mass were considered: 100, 300 and 1000 GeV.  Although branching ratios of WIMPs to photons are needed to predict event rates, we chose to be agnostic of this value and simply acknowledge that the branching ratio is generally quite small.   
\paragraph*{}
Figure 1 shows the fraction of solar captured WIMP annihilations which occur outside of the Sun as a function of WIMP-WIMP cross section for a few WIMP masses.  The results indicate that to observe gamma rays from near-solar WIMP annihilations, an experimental method about 16 orders of magnitude more effective than those experiments designed to detect internal solar decay products would be needed.  Neutrino telescopes have approximately the following probability of detecting a neutrino passing though it's effective area$^5$:

\begin{equation}
Probability=2 \times 10^{-13} m^{2}_{WIMP} (GeV)
\end{equation}

Therefore, a gamma ray detector with similar effective area and perfect efficiency would still observe many orders of magnitude fewer events than neutrino telescopes.  The most optimistic neutrino event rates from WIMP annihilation are generally of the order of tens or a hundred events per year for a square kilometer effective area$^2$.  
\paragraph*{}
To consider the most optimistic case, we can describe a truly enormous gamma ray detector of, for example, 100 square kilometers effective area, which is perfectly able to differentiate between background and actual signal from near-solar WIMPs (the Veritas detector, which isthe largest such detector planned, is about 3 orders of magnitude smaller and perfect efficiency is not actually realistic).  Suppose that the number of gamma rays produced in such annihilations is of the same order of magnitude of the number of neutrinos produced.  This number may reach up to the order of $10^{12}$ events per square kilometer of detector area per year in the most favorable supersymmetric models (generally the number is far fewer)$^2$.  Then, together, such a scenario would yield less than one event per hundred years.  This result neglects the difference of branching ratios of WIMPs to photons and neutrinos.  This effect would make the gamma ray event rates even lower. 
\paragraph*{}  
The existing and planned gamma ray detectors include large area Earth-based detectors such as MILAGRO and small area space-based detectors such as GLAST.  GLAST has the advantage of energy sensitivity.  WIMP annihilation to a photon pair would create a spectral line at the mass of the WIMP.  The energy sensitivity of such as detector could be used to remove background events, but still suffers from the lack of signal described above.  
\paragraph*{}
Another class of solar-captured WIMPs has been proposed.  WIMPs in this class have had their orbits modified by planetary gravitational perturbations such that their orbits no longer intersect the sun, and thus are stable$^6$.  Although this population may be able to enhance the near-solar WIMP density and thus the near-solar annihilation rates, the enhancement is generally less than a factor of two and not particularly relevant for this discussion.
\paragraph*{}

\begin{figure*}[t]
\begin{center}
  \epsfig{file=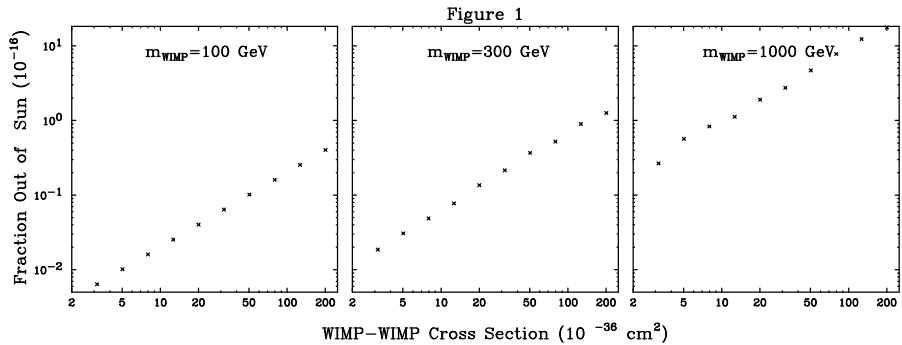,width=12cm} 
\end{center}
\vspace*{-3mm}
\begin{center}
  \caption{Fraction of solar WIMP-WIMP annihilations which occur outside of the sun for three cases of WIMP mass as a function of WIMP-WIMP annihilation cross section.  Note the factor of $10^{-16}$ included in the vertical axix.}
\end{center}
\end{figure*} 

\LARGE{IV. Conclusions}\
\setlength{\parindent}{0pt}

\normalsize{}
\paragraph*{}
The very small number of solar captured WIMP annihilations which occur outside of the sun makes the prospect of WIMP-produced gamma ray detection very poor indeed.  Detectors such as MILAGRO would need very unrealistic effective areas to compete with existing and future neutrino telescopes in this type of observation.  
\paragraph*{}

\LARGE{Acknowledgements}\
\setlength{\parindent}{0pt}

\normalsize{}
\paragraph*{}
Thanks to Brenda Dingus and Francis Halzen for valuable discussions.  This research was supported in part by DOE grant No. DE-FG02-95ER40896 and in part by the Wisconsin Alumni Research Foundation.
\paragraph*{}

\LARGE{References}\
\setlength{\parindent}{0pt}

\normalsize{}
\vspace{5mm}
[1]L.~Bergstrom,
Rept.\ Prog.\ Phys.\ {\bf 63}, 793 (2000)
[hep-ph/0002126].

\vspace{5mm}
[2] V. Barger, F. Halzen, D.W. Hooper, C. Kao, MADPH-00-1195, {\it to be published.}

\vspace{5mm}
[3] S.~C.~Strausz,
{\it Prepared for 26th International Cosmic Ray Conference (ICRC 99), Salt Lake City, Utah, 17-25 Aug 1999}.

\vspace{5mm}
[4] G.~Jungman, M.~Kamionkowski and K.~Griest,
Phys.\ Rept.\ {\bf 267}, 195 (1996)
[hep-ph/9506380].

\vspace{5mm}
[5] F.~Halzen, {\it Prepared for International Symposium on Particle Theory and Phenomenology, Iowa State University, 22-24 May 1995}, astro-ph/9508020.

\vspace{5mm}
[6] L.~Bergstrom, T.~Damour, J.~Edsjo, L.~M.~Krauss and P.~Ullio, JHEP{\bf 9908}, 010 (1999) [hep-ph/9905446].

\end{document}